\documentclass{article}
\usepackage{ijcai23}

\usepackage{times}
\usepackage{soul}
\usepackage{url}
\usepackage[hidelinks]{hyperref}
\usepackage[utf8]{inputenc}
\usepackage[small]{caption}
\usepackage{graphicx}
\usepackage{amsmath}
\usepackage{amsthm}
\usepackage{booktabs}
\usepackage{algorithm}
\usepackage{algorithmic}
\usepackage[switch]{lineno}
\usepackage{booktabs}

\usepackage{amssymb}
\usepackage{multirow} 
\usepackage{booktabs}
\usepackage{makecell}
\usepackage{color}
\usepackage{float}
\usepackage{graphicx}
\graphicspath{{./figures/}}
\usepackage{bm}
\usepackage{setspace}
\usepackage{array}
\usepackage{amsthm}
\theoremstyle{plain}

\theoremstyle{definition}

\theoremstyle{remark}

\usepackage{bm}

\newtheorem{theorem}{Theorem}

\title{Sequential Attention  Source Identification Based  on Feature Representation}

\author{
Dongpeng Hou$^{1,2}$
\and
Zhen Wang$^{1,2*}$\and
Chao Gao$^{2*}$\And
Xuelong Li$^2$\\
\affiliations
$^1$School of Mechanical Engineering, Northwestern Polytechnical University (NWPU)\\
$^2$School of Artificial Intelligence, Optics and Electronics (iOPEN), Northwestern Polytechnical University (NWPU)\\
\emails
w-zhen@nwpu.edu.cn, 
cgao@nwpu.edu.cn
}

\begin{document}

\maketitle

\begin{abstract}

Snapshot observation based source localization has been widely studied due to its accessibility and low cost. However, the interaction of users in existing methods does not be addressed in time-varying infection scenarios. So these methods have a decreased accuracy in heterogeneous interaction scenarios. To solve this critical issue, this paper proposes a sequence-to-sequence based localization framework called \textbf{T}emporal-sequence based \textbf{G}raph \textbf{A}ttention \textbf{S}ource \textbf{I}dentification (TGASI) based on an inductive learning idea. More specifically, the encoder focuses on generating multiple features by estimating the influence probability between two users, and the decoder distinguishes the importance of prediction sources in different timestamps by a designed temporal attention mechanism. It's worth mentioning that the inductive learning idea ensures that TGASI can detect the sources in new scenarios without knowing other prior knowledge, which proves the scalability of TGASI. Comprehensive experiments with the SOTA methods demonstrate the higher detection performance and scalability in different scenarios of TGASI.

\end{abstract}

\section{Introduction}
Numerous studies have demonstrated that rumors spread quickly in social networks~\cite{qian:2018neural,wang:2022rapid}, which could be extremely harmful to society. The issues of automatically resisting rumors, i.e., content-based rumor identification and rumor source identification,
have garnered much attention in recent years~\cite{dong:cikm,jin:2017multimodal,song:2021multimodal,lao:2021rumor,yang:2021rumor,khoo:2020interpretable}.   
However, it is still difficult to effectively cut off and control the propagation of rumors in social networks if only to identify the fake news instead of locating the rumor sources. 
Therefore, more and more studies focus on the problem of rumor source identification in social networks based on the network structure~\cite{jiang:2016identifying,jin:2021schemes}.

\begin{figure}[!ht]
	\centering
	\includegraphics[width=3.4in]{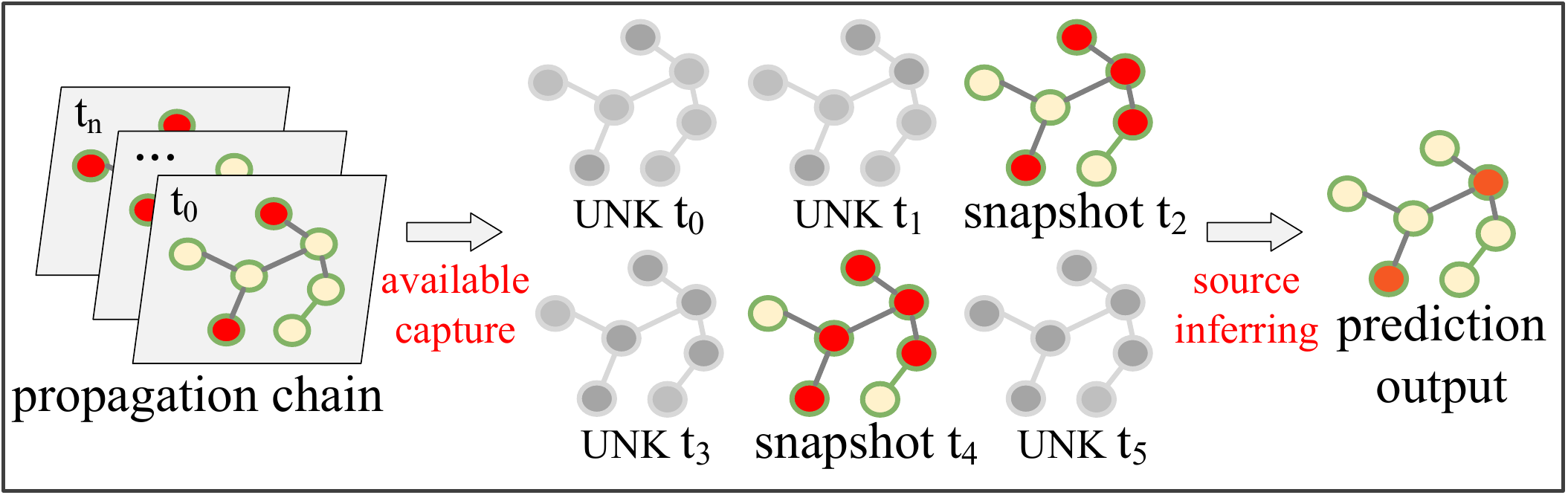}
	\caption{The illustration of the multiple rumor source detection problem under available snapshots with discrete timestamps. Due to the constraints of time and space consumption, the timestamps of snapshots we can capture are limited and discrete.}
	\label{fig1}
\end{figure}

The corresponding methods of rumor source identification are classified into two types, including infection-status-based methods and sensor-based methods~\cite{dong:cikm}.
In practice, the sensors sometimes cannot be presciently deployed in the break area in advance, and the sensor deployment process requires time and space overhead~\cite{paluch:2020optimizing}. 
In contrast, the infection-status-based methods have broader application scenarios because the snapshots are easier to obtain in practice~\cite{diepold:snapshots2022data}.

However, most of the infection-status-based methods, e.g., IVGD~\cite{wang:2022www}, SL\_VAE~\cite{ling:2022kdd}, assume that the attributes of all individuals (e.g., the infection probability or influence intensity) in a network is known in advance.
Obviously, it is difficult to obtain such  information in practice due to the huge labor and statistical costs. 
Although the SOTA methods, e.g., GCNSI~\cite{dong:cikm} and GCSSI~\cite{dong:2022trans}, try to identify the rumor sources without using the information of the underlying propagation model,
they do not explicitly consider the user's interactive behavior when several time-varying   snapshots are conveniently available. 
Therefore, the lack of essential behavior representation  results in the limited performance of current localization methods in the complicated social network.
Besides, the scalability of the model reflects the extensive application and robustness in practice~\cite{agarwal:2021transfer-inductive}.
But the source localization methods so far have not considered the transfer performance of the model in new scenarios (different propagation models or different social networks, denoted as inductive learning~\cite{michalski:1983theory}).
In summary, there is no sequence based inductive  localization model for time-varying snapshots to  consider the impact of user's behavior features on propagation.

In this paper, we study the multiple rumor source detection (MRSD) problem and focus on addressing the above issues.
Our goal is to design a generic source localization framework with transferability, especially can analyze the  behavior diversity in time-varying infection scenarios.
In practice, there should be a potentially heterogeneous influence between any two users in a social network no matter what the complex propagation models are. 
And such the corresponding influence can be evaluated from a collection of available  snapshots with time-varying infection characteristics in discontinuous timestamps.
So we can infer the sources by using these available captured snapshots.
Such a problem is presented in Fig.~\ref{fig1}. 
To this end, considering the excellent embedding ability of Graph Neural Network (GNN)~\cite{welling:2016GCNICLR} for node attributes (or node features) and the outstanding modeling ability of the Gate Recurrent Unit (GRU)~\cite{chung:2014empiricalGRU} for time series data, we propose a novel sequence-to-sequence model called Temporal-sequence based Graph Attention Source Identification (TGASI).
The input is a set of infection snapshots at available timestamps in a social network, and the output is a set of predicted sources.
The TGASI model includes a GNN-based encoder to generate and aggregate the low-dimensional feature embedding of each individual, and a GRU-based decoder to infer the source with a temporal attention mechanism.
The major contributions of TGASI are as follows.
\begin{itemize}
\item
A generic sequence-to-sequence framework with transferability is designed  for source localization, enabling TGASI to transfer to new diffusion models and social networks for source prediction without knowing other prior knowledge.
And a unique loss function is designed to further enhance the performance of TGASI in the source localization task. 

\item
We design a deep module based on the stationary distribution of a Markov chain to learn the user's probability transition matrix in order to differentiate the user behavior diversity and consider the influence of heterogenous behavior on propagation. Then a high quality representation of the interactive behavior embedding is obtained.

\item
We design the dynamic infection  features and static topology features to enrich the representation of low-dimensional embedding. The propagation information based dynamic features can better fit the source localization task, and topological features can better incorporate the graph-level  task into a sequence based framework.

\item
A one-timestamp based attention mechanism is introduced to further comprehensively consider the weight of the source prediction probability at different timestamps. Dynamically distinguishing the importance of different timestamps can solve the imbalance of infection information under temporal characteristics.

\end{itemize}

\section{Related Work} \label{sec02}
TGASI is a sequence based framework to solve the source localization problem based on the infection status observations. Therefore, the infection-status-based localization and sequence-to-sequence learning are summarized.
\subsection{Infection-status-based Source Localization}
The aim of the source localization is to infer a set of sources in a network based on some observations.
Due to the convenience and feasibility of snapshot acquisition, 
several recent surveys on this topic are available~\cite{jiang:2016identifying,shelke:2019source}.
A novelty method  LPSI designs a source prominence based label propagation method to identify the source without using the prior knowledge~\cite{wang:2017multipleAAAI}.
With the development of deep learning, Dong et al. inspired by LPSI design a GCN based source identification (GCNSI) model to solve the MRSD problem~\cite{dong:cikm}.
However, they only simply consider the centrality features of individuals.
Therefore, some studies focus on more features in the source localization task, such as the SIGN method extracting the individual's dynamic features~\cite{li:2021JBI}, and the MCGNN method considering the edge features~\cite{sun:2021multichanel}.
Moreover, the dynamic features of the propagation are also considered before executing the source inferring process, like IVGD~\cite{wang:2022www}, SL\_VAE~\cite{ling:2022kdd}. 
However, all the above methods do not consider and analyze the impact of time series characteristics on source localization.
In summary, there is no generic sequence-to-sequence model for source localization to comprehensively consider the behavior features under available time-varying snapshots.

\subsection{Sequence-to-Sequence Learning}
With the rapid development of deep learning, sequence-to-sequence learning shows great performance on various tasks based on the characteristic of time series ~\cite{sutskever:2014sequence}. For example, 
Chung et al. propose the gated recurrent unit (GRU), whose performance is comparable to LSTM~\cite{hochreiter:1997longLSTM} with lightweight overhead~\cite{chung:2014empiricalGRU}. And GCSSI is a sequence-to-sequence framework using GRU as a recurrence unit to learn the time series features~\cite{dong:2022trans}. Currently, the novel encoder-decoder architecture for supervised discrete time dynamic graph (DTDG) learning focuses on the various dynamics in the real world, which brings new ideas to the embedding of temporal information~\cite{zhu:2022learnableDynamic}. TGASI is based on the one of DTDG architectures, called the static graph encoder and sequential decoder, for the MRSD task.
The dynamic intuition in TGASI refers to the dynamic infection features of each individual in different timestamps. 
We improve such a framework to analyze the  behavioral features of individuals under time-varying characteristics based on the inductive learning idea, so that the trained TGASI can be applied to different networks and different propagations. 

\begin{figure}[htb]
	\centering
	\includegraphics[width=2.95in]{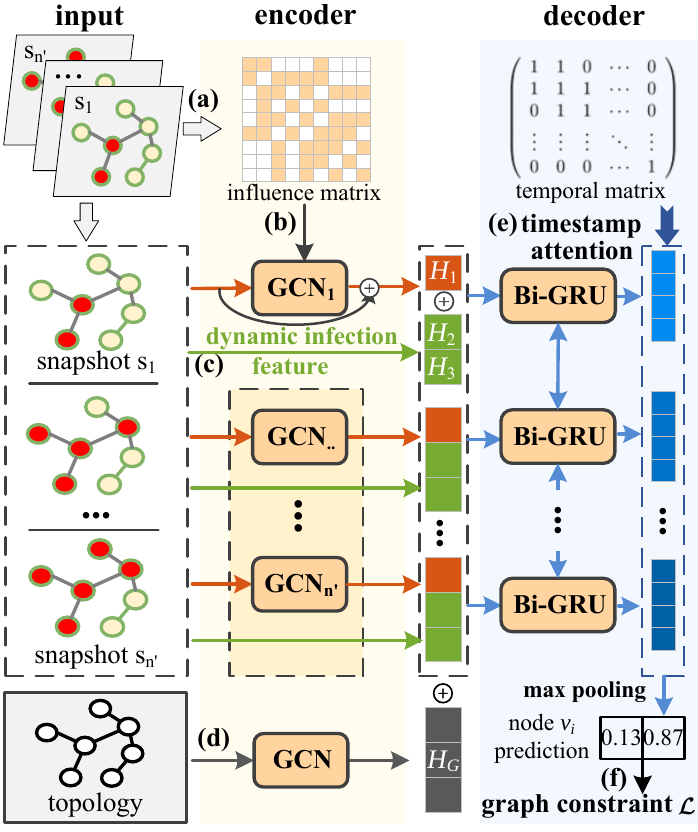}
	\caption{The illustration of TGASI based on the sequence-to-sequence framework.
	The features generation encoder includes four innovative  modules. (a) influence probability transition matrix $\bm{\mathcal{W}}$ is estimated by available captured snapshots.
	(b) coarse-grained source probability feature in one timestamp $s$ is designed based on the influence matrix $\bm{\mathcal{W}}$.
	(c) dynamic infected feature $H_2$ and uninfected feature $H_3$ is generated from each node's neighbors.
	(d) lower dimensional embedding $H_G$ of the topology structure is obtained based on the topology graph. 
	The GRU-based decoder with temporal attention uses Bi-GRU to decode the embedding, then (e) a one-timestamp based attention mechanism is designed for time series information in order to distinguish and weight the importance of decoder information at each different timestamp for source localization task.
	What's more, (f) a graph constraint based loss function is specially designed for the localization task to train TGASI.
	}
	\label{figframework}
\end{figure}

\section{Methodology}\label{sec03}

\subsection{Problem Definition}
Given a collection of captured available snapshots 
$\{ \mathcal{G}_{s_j}=(V, E,Y_{s_j}) \mid j=1,~2,~3,... \}$ with different timestamps $s_j \in T=\{t_1,~t_2,~t_3,...\}$ based on an undirected  network $G$ = $(V, E)$, where $T$ is the timestamp set of complete propagation chain (including unavailable snapshots, i.e., $t_1$$\leq$$s_1$\textless$s_{n'}$$\leq$$t_n$, and $n'$\textless$n$),  $Y = \{Y_1, ...,Y_{|V|}\}$ is the set of the node state, and  $Y_i$=1 if the observed state of node $v_i$ is infected, otherwise $Y_i$=0. 
And we denote the original rumor sources set as $R$ $\subset$ $V$. The goal of our method in the MRSD problem is to predict a source set $R^{*}$ which can maximize indicator like $\frac{R \cap R^{*}}{R \cup R^{*}}$.

\subsection{Temporal-sequence Based Graph Attention Source Identification }\label{3.2}

In this part, an inductive framework based on the graph-level task  for  source localization is proposed, namely temporal-sequence 
based graph attention source identification.
As shown in Fig.~\ref{figframework}, the framework of TGASI  is an encoder-decoder essentially. More specifically, a GNN-based encoder generates and embeds the node's features, which include coarse-grained-based source probability, dynamic infection  features, and static topology features. And a GRU-based decoder  with a temporal attention mechanism infers the source by decoding the embedding from the encoder. 

\subsubsection{Features Generation Encoder}

Considering the complicated behavior diversity in social networks in practice, and such the potential relationship of any two users can be characterized as a heterogenous influence.
Then a deep heterogeneous influence estimation method is proposed to automatically learn the influence matrix $\bm{\mathcal{W}} \in \mathbb{R}^{|V| \times |V|}$,
where $\mathcal{W}_{ij}$ represents the probability that an infected-state node $v_i$ successfully influences its uninfected neighbor $v_j$. Note that  $\mathcal{W}_{ij}$$\neq$$\mathcal{W}_{ji}$ because people are influenced by celebrities to buy endorsement products but celebrities do not be influenced in reverse.
More specifically, a propagation mechanism $f_\theta(\mathcal{G}_{s_j})$ is used to predict the infection state of the snapshot of the next timestamp $\mathcal{G}_{s_{j+1}}$.
Here the $f_\theta$ is a nonlinear mapping function to characterize a  propagation scheme based on the learnable matrix $\bm{\mathcal{W}}$. 
Based on the mapping, we can construct the predicted snapshots matrix $\mathcal{G'}$, i.e.,
\begin{equation}
\label{eq1}
\mathcal{G'} = [f_\theta(\mathcal{G}_{s_1}), f_\theta(\mathcal{G}_{s_2}),
f_\theta(\mathcal{G}_{s_3}),...]
\end{equation}

Note that the selection of the propagation mapping function $f_\theta$ is flexible, and it can be designed autonomously based on deep models, e.g., GCNs~\cite{welling:2016GCNICLR} or attention-based graph models~\cite{velivckovic:2018GAT}. 
As for the detailed underlying design, 
some research has demonstrated that users in social networks have the ability to identify and prevent fraud~\cite{baesens:2015fraud,ventola:2014social}.
So we use the IC strategy that only has one opportunity to activate the neighbors instead of the common widely used random walk that can be infected multiple times. 
What's more, the stationary distribution of a Markov chain  revises the distribution bias of influence parameters (i.e., $\bm{\mathcal{W}}$) in the diffusion process~\cite{salamat:2020balnode2vec}. 
Motivated by the stationary distribution of influence diffusion models~\cite{xia:WSDM2021deepis}, an approximate stationary distribution of a Markov chain based on the IC diffusion model is adopted for $f_\theta$, as defined in Eq.~(\ref{eq4}).
\begin{equation}
\label{eq4}
f_\theta(Y_{v_i}^{s+1})=1-\prod_{v_j \in \mathcal{N}\left(v_i\right)}\left(1-\mathcal{W}_{j i} 
\left(Y_{v_j}^{s}-Y_{v_j}^{s-1} \right) \right)
\end{equation}
where $\mathcal{N}(v_i)$ is the neighbors set of node $v_i$, and $Y^s \in \mathbb{R}^{|V|}$ is an available snapshot in timestamp $s$.
The equation means that the state of node $v_i$ in the next timestamp is approximately influenced by the heterogenous behavior of its neighbors in the timestamp $s$. In order to further improve the efficiency of the learning process of this propagation parameter on the graph-level task, we manually design two rules that conform to propagation. First, we set $\mathcal{W}_{ij}$=0 and remove the corresponding learning gradient if there is no edge between $v_i$ and $v_j$.  Second, the propagation parameter theoretically should be greater than 0, and we take the absolute value when the learning parameter $\mathcal{W}_{ij}$$<$$0$ in order to increase the penalty of MSE loss.
Then we generalize the nonlinear mapping function $f_\theta$ from one node to network $G$, 
and use $f_\theta(\mathcal{G})$ to guide an MLP to  evaluate the influence matrix  $\bm{\mathcal{W}}$.

Although there are many classic sequence-to-sequence models such as AGC-Seq2Seq~\cite{zhang:2019trafficTransfer} and  i-Revnet~\cite{jacobsen:2018revnet,chang:2018reversible} to directly identify the source based on the data of the snapshots,  the sparse features (we only have two available features whether the state is infected or not) and lack of parameters in hidden layers restrict the ability to learn the spatial-temporal information in the localization task. So we construct some necessary feature embedding based on user behavior diversity $\bm{\mathcal{W}}$.

First, We consider the coarse-grained source probability based on the behavioral interaction under each independent snapshot $s$, which is denoted as $H_1^s$.
GNN can efficiently and conveniently embed the user's influence information in $\bm{\mathcal{W}}$ into the message-passing process.
Inspired by the invertible graph residual network, $H_1^s$ is designed by using the following graph convolution operation.
\begin{equation}
\label{eq6}
\begin{split}
H_1^{s}&=\varepsilon^{s}_{\mathrm{GRN}}(Y^{s}, \boldsymbol{A} \mid \bm{\mathcal{W}})=
\varepsilon^{s}_{\mathrm{GCN}}(Y^{s}, \boldsymbol{A} \mid \bm{\mathcal{W}}) + \alpha Y^{s} \\
&=\sigma\left(\widehat{\boldsymbol{D}}^{-\frac{1}{2}} \widehat{\boldsymbol{A}} \widehat{\boldsymbol{D}}^{-\frac{1}{2}} \boldsymbol{X} \boldsymbol{W}\right) + Y^{s}, 
s.t.~L_\varepsilon < 1, \\
&\widehat{\boldsymbol{A}} =  \boldsymbol{A}+ \boldsymbol{I_n}+ \bm{\mathcal{W}}, 
X_i= \begin{cases}{[1,0],} & Y_{v_i}^s=0 \\ {[0,1],} & Y_{v_i}^s=1\end{cases}
\end{split}
\end{equation}
where $\bm{W} \in \mathbb{R}^{2 \times h_{one}}$ is a learnable weight matrix in the GCN module, $2$ is the number of observed infection features which imply the infected state and uninfected state (denoted as $X_i$ here), $h_{one}$ is the hidden size of GCN in one timestamp $s$, $\sigma(\cdot)$ is the activation function, $H_1^{s}$$\in$$\mathbb{R}^{|v|}$ is the output feature of all nodes in a timestamp $s$, $\alpha$ is the residual weight, and $L_\varepsilon$ is the  Lipschitz constants  to initialize $\bm{W}$. Here, we convert the observed state feature $Y$ of snapshots into a one-hot encoding, characterizing that the infected state and uninfected state are categorical variables rather than numeric variables. The operation can enhance the learning ability and interpretability of the deep module.
What's more, the learnable weight matrix $\bm{W}$ is normalized via the power iteration method to achieve the Lipchitz constant $L_\varepsilon$$<$$1$ in order to guarantee the stronger robustness and accelerate  convergence speed~\cite{wang:2022www}. 
Then, we consider the infection characteristics of all nodes in each available timestamp $s$ according to the source localization task, and design two infection features of node $v_i$ in timestamp $s$, which is shown in Eq.~(\ref{eq7}) and Eq.~(\ref{eq8}).
\begin{equation}
\label{eq7}
H_2^{s}(v_i) = \frac{\sum_{v_j \in \mathcal{N}\left(v_i\right), Y_{j}^{s}=1}\mathcal{W}_{ij}}{|\mathcal{N}\left(v_i\right)|}
\end{equation}

\begin{equation}
\label{eq8}
H_3^{s}(v_i) = \frac{\sum_{v_j \in \mathcal{N}\left(v_i\right), Y_{j}^{s}=0}\mathcal{W}_{ij}}{|\mathcal{N}\left(v_i\right)|}
\end{equation}
where $H_2^{s} \in \mathbb{R}^{|V|}$ and $H_2^{s}(v_i)$ is a infected neighbor's feature of $v_i$ in an available timestamp $s$, and similarly $H_3$ is the uninfected neighbor's feature. 
The $H_2$ and $H_3$ are interpretable, and the proof is in the appendix due to the limited space.

\begin{theorem}
    The ratio of the neighbor's infected state and uninfected state are special variants of infected feature $H_2$ and uninfected feature $H_3$, respectively.
\end{theorem}

The aim of designing these embedding (i.e., $H_1$-$H_3$) is to make the decoder efficiently learn the propagation features and predict the source. However, the common input of the time series decoder is sequence based data. It is necessary to additionally design the rules if the input includes graph information.
There are two reasons why we do not directly process the decoder in the form of graph recurrent layer, such as $\bm{h_t}=\sigma \Big(\bm{A} \bm{H^{s}} \bm{W_{i h}}+\bm{A} \bm{h_{t-1}} \bm{W_{h h}}+b \Big) $.
First, the format constraints of the recurrent unit $\bm{W_{ih}}$$\in$$\mathbb{R}^{3 \times 2}$ and $\bm{W_{hh}}$$\in$$\mathbb{R}^{2 \times 2}$ would   lead to too few learnable parameters and reduce the model's learning capacity. The second reason is that designing topological features in the encoder expands the dimensionality of the embedding and increases the quality and interpretability of the embedding.
So we use a one-layer GCN to design the topology feature embedding in a low-dimensional way to better solve the graph-level based localization task, as shown in Eq.~(\ref{eq10}).

\begin{equation}
\label{eq10}
\begin{split}
H_{G} &= \varepsilon_{\mathrm{GCN}_{G}}(\bm{I_n}, \boldsymbol{A} ) =\sigma\left(\widehat{\boldsymbol{D}}^{-\frac{1}{2}} \widehat{\boldsymbol{A}} \widehat{\boldsymbol{D}}^{-\frac{1}{2}} \bm{I_n} \boldsymbol{W}\right) 
\end{split}
\end{equation}
where $\bm{W} \in \mathbb{R}^{|V| \times \sqrt{|V|}}$ is a learnable weight matrix, and $H_G \in \mathbb{R}^{|V| \times \sqrt{|V|}}$ is the output topology features of all nodes. 
Therefore, we implement the feature embedding from a high-dimensional graph to a low-dimensional vector.
 
We get the coarse-grained source probability $H_1^{s}$, the infection features $H_2$ and $H_3$ in the timestamp $s$, and the network topology $H_G$ from the encoder. And the embedding $H$ in the timestamp $s$ can be obtained by concatenating these features.
\begin{equation}
\label{eq11}
H^{s} = \mathrm{CONCAT} (H_1^{s}, H_2^{s}, H_3^{s}, H_G)
\end{equation}

\subsubsection{GRU-based Decoder with Temporal Attention}
For the available snapshots 
$\{ \mathcal{G}_{s_j}=(V, E,Y_{s_j}) \mid j=1,~2,~3,... \}$ with different timestamps $s_j$, we can acquire the embedding $H^{s_1}$, $H^{s_2}$, $H^{s_3}$, etc., from the encoder. Next, a time series  decoder needs to be designed to predict the source by using the embedding. To determine the output dimension,
we identify the source localization problem as a node binary classification problem, 
then we use a target function that includes the loss of all nodes to constrain the training on graph-level $G$.
A bidirectional lightweight time series module, i.e., Bi-GRU, is used as the decoder to predict the source.
The update function is shown in Eq.~(\ref{GRU}).
\begin{equation}
\label{GRU}
h_t=\left(1-z_t\right) * n_t+z_t * h_{(t-1)} 
\end{equation}
where 
$h_{t}$ and $h_{t-1}$ are the hidden layer state of the model at the timestamp $s_t$ and the previous timestamp, 
$z_t$ and $n_t$ are the update and new gates, respectively.
We input the embedding of snapshot sequence $\left\{H^{s_1},H^{s_2},H^{s_3},...H^{s_{\zeta}}\right\}$ into a Bi-GRU model, and the forward GRU infer the source based on the snapshots from the timestamp $s_{1}$ to $s_{\zeta}$, and we denote the output predicted binary classification for all nodes of the hidden layer in the timestamp $s_j$ as $\overset{\rightarrow}{h}_{j} \in \mathbb{R}^{|V| \times 2}$. 
Similarly, the output of backward GRU
is represented as  $\overset{\leftarrow}{h}_{j} \in \mathbb{R}^{|V| \times 2}$. Finally, the two hidden layer states of the timestamp $s_j$ are concatenated, i.e.,  
$\hat{R}_{j}=\left[\overset{\rightarrow}{h}_{j} , \overset{\leftarrow}{h}_{j} \right] \in \mathbb{R}^{|V| \times 4}$. 

For any timestamp $s_j$, each node $\hat{R}_{j}(v)$ has $4$ features output by the Bi-GRU. 
However, it is difficult to determine exactly at which timestamp the source probability can be evaluated. Because the infection information is imbalanced in different timestamps. 
That is, the information in the earlier propagation stage is lacking. 
And the information in the later propagation stage is highly coupled and overlapping.
Thus, we design a temporal attention mechanism to dynamically adjust the  weight of predicted source probability in different timestamps. 
A one-timestamp adjacency matrix $\bm{\varpi}$ is designed to enhance the attention ability for the infection timestamps, which is defined as follows.

\begin{equation}
\label{eq17}
\bm{\varpi}=\left(
\begin{array}{ccccc}
1 & 1 & 0 & \cdots & 0 \\
1 & 1 & 1 & \cdots & 0 \\
0 & 1 & 1 & \cdots &0 \\
\vdots & \vdots & \vdots & \ddots & \vdots \\
0 & 0 & 0 & \cdots & 1
\end{array}
\right)
\end{equation}

And the self-attention focusing on the weight of different  timestamps is implemented by a shared attentional mechanism $a$: 
$\mathrm{CONCAT}(\mathbb{R}^{|\bm{W_A}^{T}\hat{R}(v)|}, \mathbb{R}^{|\bm{W_A}^{T}\hat{R}(v)|})$$\rightarrow$$\mathbb{R}$ to compute the attention coefficients $e_{ij}$ corresponding to the timestamp $s_i$ and timestamp $s_j$ of a node $v$, which indicates the importance of timestamp $j$’s features to timestamp $i$.
\begin{equation}
\label{eq18}
e_{ij}(v) = a(\bm{W_A}^{T}\hat{R}_i(v), \bm{W_A}^{T}\hat{R}_{j}(v)) \quad s.t. \quad \mathrm{abs}(i - j) \leq 1
\end{equation}
where $\bm{W_A} \in \mathbb{R}^{4 \times 2}$,
here the input feature $4$ is the double-times binary classification.
And we apply a single-layer BP neural  network for the  attention mechanism $a$, which is a weight vector with $\vec{a} \in \mathbb{R}^{2|\bm{W_A}^{T}\hat{R}(v)|}$. 
Then to make coefficients easily comparable across different timestamps, we normalize the attentional mechanism.
\begin{equation}
\small
\label{eq19}
\begin{gathered}
\phi_{i j}(v)=\operatorname{softmax}_j\left(e_{i j}(v)\right)=\frac{\exp \left(e_{i j}(v)\right)}{\sum_{k \in \bm{\varpi}_{i k}} \exp \left(e_{i k}(v)\right)} \\
=\frac{\exp \left(\operatorname{ReLU}\left(\overrightarrow{\mathbf{a}}^T\left[\bm{W_A}^{T}\hat{R}_i(v) \|  \bm{W_A}^{T}\hat{R}_{j}(v)\right]\right)\right)}{\sum_{k \in \bm{\varpi}_{i k}} \exp \left(\operatorname{ReLU}\left(\overrightarrow{\mathbf{a}}^T\left[\bm{W_A}^{T}\hat{R}_i(v) \|  \bm{W_A}^{T}\hat{R}_{j}(v)\right]\right)\right)} \\
\end{gathered}
\end{equation}
The normalized attention coefficients $\phi_{i j}$ are used to compute a linear combination of the four features in one timestamp corresponding to them, to serve as the final source prediction for a node $v$ in timestamp $s_i$ with $\mathcal{K}$-head attention, which is shown in  Eq.~(\ref{eq20}).

\begin{equation}
\label{eq20}
\begin{gathered}
\hat{R}_{i}^{\prime}(v)
=\mathrm{softmax}\left(\frac{1}{\mathcal{K}} \sum_{1}^\mathcal{K} \sigma\left(\sum_{j \in \bm{\varpi}_{i j}} \phi_{i j} \bm{W_A}^{T}\hat{R}_i(v)\right)\right) \\
\end{gathered}
\end{equation}

After the final prediction $\hat{R}_{i}^{\prime}(v)$ for each node $v$ in each timestamp $s_i$ finishes, we pick the needed number (denoted as $Z$) of predicted nodes by ranking the highest  probability as the predicted source, such a source set $R^{*}$ is shown in Eq.~(\ref{eq21}).
\begin{equation}
\label{eq21}
R^{*}=\{\underset{v \in V }{\operatorname{argmax}}~( \hat{R}_{i}^{\prime}(v)) \mid v \notin R^{*} , |R^{*}|=Z\}
\end{equation}

\subsection{Loss Function}
Further, 
we design a unique loss function for TGASI to better adapt to the task of source localization.
\begin{equation}
\mathcal{L}=\mathcal{L}_{\textit{Entropy}}(R_{\mathrm{hot}},R^{*}_{\mathrm{hot}})+\mathcal{L}_{\textit{MSE}}(R_{\mathrm{hot}},H_1^{s_i})+\mathcal{L}_{G}(R_{\mathrm{hot}},R^{*}_{\mathrm{hot}})
\end{equation}

Among them, $\mathcal{L}_{Entropy}$ using binary cross entropy is the main loss to train the complete process of TGASI.
$\mathcal{L}_{MSE}$ using mean square error is an encoder loss to constrain the learning process of source probability prediction for a single timestamp $s_i$. 
And $\mathcal{L}_{G}$ is a designed auxiliary loss to enhance the learning ability of the TGASI on the localization task.

\begin{equation}
\begin{split}
\label{eq25}
\mathcal{L}_{G}(R_{\mathrm{hot}},R^{*}_{\mathrm{hot}})= 
\frac{\sum R_{\mathrm{hot}}}{|V|}
\sum_{v_i \in R_{\mathrm{hot}}(v_i)=0}(R^{*}_{\mathrm{hot}}[v_i,0])\\
+(1-\frac{\sum R_{\mathrm{hot}}}{|V|})
\sum_{v_i \in R_{\mathrm{hot}}(v_i)=1}(R^{*}_{\mathrm{hot}}[v_i,1])
\end{split}
\end{equation}

\section{Experiments}\label{sec04}

\subsection{Datasets and Baselines}
Six social networks are selected to evaluate the performance of all localization methods~\footnote{The six datasets are available at: http://snap.stanford.edu}.
\begin{table}[htb] 	
	\renewcommand{\arraystretch}{1.1}
	\label{table_network}
	\centering
	\setlength{\tabcolsep}{3.5mm}{
	\begin{tabular}{cccccc}
		\toprule	
		 & Network & $|V|$ & $|E|$ & $\langle$k$\rangle$ \\
		\midrule
        $G_1$ & Karate & 34 & 78 & 4.6 \\
		$G_2$ & Jazz & 198 & 2742 & 27.7 \\
		$G_3$ & Facebook& 4039 & 88234 & 43.69 \\
		$G_4$ & Twitch-ES & 4648 & 59382 & 25.55  \\
		$G_5$ & Wiki-Vote & 7115 & 103689 & 29.15 \\
        $G_6$ & Page-Large & 22470 & 171002 & 15.22  \\
		\bottomrule
	\end{tabular}}
 \caption{Information of the real-world datasets. 
	}
\end{table}

To demonstrate the validity and novelty of the current work, we compare the TGASI method with the SOTA localization methods, including IVGD~\cite{wang:2022www}, SL\_VAE~\cite{ling:2022kdd}, GCSSI~\cite{dong:2022trans}, SIGN~\cite{li:2021JBI}, MCGNN~\cite{sun:2021multichanel}, ResGCN~\cite{shah:2020patient0}, GCNSI~\cite{dong:cikm}.

\subsection{Evaluation Metrics}
By comprehensively considering the evaluation metrics of the baseline methods, we use two metrics, i.e., the standard F1-score~\cite{sokolova:2006F1} (F1) and average error distance (AED)~\cite{dong:2022trans}. 
\begin{equation}
\label{eq6.1}
F1\verb|-|score=  \frac{2 * { Precision } *  { Recall }}{ Precision +\ { Recall }}
\end{equation}
\begin{equation}
\Delta_{\mathrm{AED}}=\min _{r^{*} \in \text { permutation }(R^{*})} \sum_{i=1}^Q \frac{d\left(r^{*}_i, r_i\right)}{Q}
\end{equation}

\begin{table*}[htb]
	\renewcommand{\arraystretch}{1.1}
	\centering
	\setlength{\tabcolsep}{1.5mm}{
\begin{tabular}{cc|ccccccccccccccccc}
\hline
\hline
\multicolumn{2}{c|}{Network}   & \multicolumn{2}{c}{$G_1$} &  & \multicolumn{2}{c}{$G_2$} &  & \multicolumn{2}{c}{$G_3$} &  & \multicolumn{2}{c}{$G_4$} &  & \multicolumn{2}{c}{$G_5$} &  & \multicolumn{2}{c}{$G_6$} \\ \hline
\multicolumn{2}{c|}{Algorithm} & F1        & AED        &  & F1        & AED        &  & F1        & AED        &  & F1        & AED        &  & F1        & AED        &  & F1        & AED        \\ \hline
\multicolumn{2}{c|}{GCNSI}     & 0.117&1.78       &  & 0.05           &1.93       &  & 0.003&2.21       &  & 0.004&2.23       &  
&0.001&2.41       &  & 0&2.89       \\
\multicolumn{2}{c|}{GCSSI}     & 0.213&1.58       &  & 0.071          &1.85       &  & 0.007&2.17    &     & 0.009 &2.18                  &  &  0.002 & 2.38      &   &  0&3.11       \\
\multicolumn{2}{c|}{SIGN}      & 0.421&1.09       &  & 0.377&1.21       &  & 0.065          &1.88       &  &0.055&1.91       &  &                0.013&2.11     &  &       0&2.86       \\
\multicolumn{2}{c|}{ResGCN}    & 0.410&1.11       &  & 0.371           &1.17       &  & 0.117&1.69       &  & 0.130&1.61       &  & 0.025          &1.89       &  & 0.007&2.17       \\
\multicolumn{2}{c|}{MCGNN}     & 0.312&1.36       &  & 0.297          &1.42       &  & 0.110 & 1.87 &  & 0.123 & 1.88               &         & 0.031& 1.98      &  &   0&2.90       \\
\multicolumn{2}{c|}{IVGD}      & $\bm{0.537}$    & 0.91      &  & 0.517&0.93       &  & 0.371&1.17         &       &  0.375&1.19                       &     &0.257&1.44       &  &0.103&2.04       \\
\multicolumn{2}{c|}{SL\_VAE}   & 0.377               & 1.18     &  & 0.353  & 1.22 &        &   0.291 & 1.39      &  &                0.277&1.43       &  &    0.196& 1.73       &  &  0.012 &   2.35    \\
\multicolumn{2}{c|}{TGASI}     & 0.514&$\bm{0.89}$       &  & $\bm{0.596}$   &$\bm{0.72}$       &  & $\bm{0.727}$   &$\bm{0.58}$       &  & $\bm{0.672}$   & $\bm{0.64}$      &  & $\bm{0.661}$   &  $\bm{0.62}$     &  & $\bm{0.493}$ & $\bm{1.49}$      \\ \hline \hline
\end{tabular}
}
\caption{Source identification performance on the test dataset of six social networks.  The bold values represent the best results.
	}
 \label{resluts}
\end{table*}

\subsection{Settings and Optimizations}\label{Optimizations}
By following existing methods in the field of information propagation~\cite{xia:WSDM2021deepis} and rumor  source localization~\cite{wang:2022www,ling:2022kdd}, we use the independent cascade (IC) model as the underlying propagation model.  
Also in reference to their study, we pick 10\% nodes as the ground-truth sources and simulate the IC propagation process based on $G_1$ to $G_6$ networks.  And we set the low infection rates in the experiments due to the propagation characteristics of rumors in social media~\cite{kou:2017supporting,wang:2022rapid}.
We independently generate 1000 sets of propagation. Each set includes several available snapshots with different timestamps.  
And we use a 10-fold cross-validation strategy to divide the training dataset and the test dataset, then the final result is output by taking the average prediction in the test dataset from each fold.
Moreover, an early stopping mechanism is designed in order to avoid over-fitting in the training process.

\subsection{ Overall Experimental Results}
The source detection performance based on the F1-score and AED metrics in $G_1$ to $G_6$ is shown in Tab.~\ref{resluts}.
The higher F1-score and the lower AED reveal better performances. 
In general, the models considering the propagation process generally have a better prediction performance than other methods, i.e., IVGD, and SL\_VAE.
And the most challenging baseline is IVGD.
The main reasons are that GCNSI, SIGN, and ResGCN only design different training models from the perspective of neural network structure, GCCSI only considers the time series features, and MCGCN only solves the single source detection rather than the MRSD task.
Although IVGD considers the individual's comprehensive interaction based on the influence, it does not consider the impact of time series characteristics in the sequence data on source detection.

Compared with the optimal baseline IVGD, TGASI is comparable in $G_1$, and TGASI based on the F1-score metric demonstrates an improvement of approximately 15\% in $G_2$, 54\% in $G_3$, 88\% in $G_4$, etc. 
Similarly, the comparison superiority of TGASI on the AED metric also increases with the network scale.
In conclusion, TGASI outperforms all the SOTA methods on the basis of rigorous metrics in all datasets. 
There are three reasons:
(1) The influence matrix is embedded through the graph residual network to imply the user interaction at each timestamp.
(2) The dynamic infection features and static topology features are designed to fit the source localization task.
(3) The attention mechanism distinguishes the weight and importance of decoder information at each different timestamp for source localization.

\subsection{Comparison of Different Loss Functions}
We specially design the loss function in the sequence-to-sequence framework to fit the source localization task.
In order to prove the necessity of  each component of the loss function $\mathcal{L}$ we designed, we rigorously demonstrate the detection  performance of TGASI by removing or replacing each item in the loss function.
Some localization studies only use MSE as the loss function, so we first use the MSE loss to replace the main loss item $\mathcal{L}_\textit{Entropy}$, which is denoted as $\mathcal{L}_{\textit{Entropy} \rightarrow \textit{MSE}}$ ($\mathcal{L}_{\textit{E} \rightarrow \textit{M}}$). 
It can be seen from Tab.~\ref{loss} that the performance of entropy loss on the TGASI model is better than that of MSE. Because we define the source localization task as a node binary classification task based on the graph-level constraint. 
What's more, the other two loss items, i.e., $\mathcal{L}_{\textit{MSE}}$ and $\mathcal{L}_{G}$ in $\mathcal{L}$, also play a positive role in the learning process of the TGASI model.
So we further keep the main loss item $\mathcal{L}_\textit{Entropy}$ unchanged, and remove the loss item $\mathcal{L}_{\textit{MSE}}$ or loss item $\mathcal{L}_{G}$, respectively.
As can be seen from the last two rows of Tab.~\ref{loss}, we can conclude that both the loss item $\mathcal{L}_{\textit{MSE}}$ and the loss item $\mathcal{L}_{\textit{MSE}}$ positively guide the source detection of TGASI.

\begin{table}[b]
\renewcommand{\arraystretch}{1.1}
	\centering
	\setlength{\tabcolsep}{0.4mm}{
\begin{tabular}{cc|cccccccc}
\hline \hline
\multicolumn{2}{c|}{Network}   & \multicolumn{2}{c}{$G_2$} &  & \multicolumn{2}{c}{$G_3$} &  & \multicolumn{2}{c}{$G_4$}  \\ \hline
\multicolumn{2}{c|}{Loss} & F1             & Epoch   &  & F1             & Epoch   &  & F1             & Epoch    \\ \hline

\multicolumn{2}{c|}{$\mathcal{L}$}     & $\bm{0.5959}$   &  5     &  & $\bm{0.7266}$   &   6    &  & $\bm{0.6716}$   &  6  \\  \hline
\multicolumn{2}{c|}{$\mathcal{L}_{\textit{E} \rightarrow \textit{M}}$}     & 0.5730   &  5     &  & 0.7013   &   6    &  & 0.6411   &  6  \\ 
\multicolumn{2}{c|}{-$\mathcal{L}_{\textit{MSE}}$}     & 0.5811   &  7     &  & 0.7132  &   6    &  & 0.6607   &  8  \\
\multicolumn{2}{c|}{-$\mathcal{L}_{\textit{G}}$}     & 0.5621  &  11     &  & 0.6933   &   10    &  & 0.6373   &  11  \\
\hline \hline
\end{tabular}
}
	\caption{The performance evaluation of variant loss in $G_2$ to $G_4$.}
	\label{loss}
\end{table}

\subsection{Ablation Study}
We further study the influence of designed components of TGASI on the source detection performance to prove their contributions. 
The critical modules of the TGASI model include the coarse-grained feature based on the estimated influence matrix $\bm{\mathcal{W}}$, the dynamic infection features $H_2$ and $H_3$, the topology features of a social network $H_G$, and the self-attention mechanism on different timestamps.
So four variant models of TGASI are developed as follows.

\begin{figure}[htb]
	\centering   
	\includegraphics[width=3.3in]{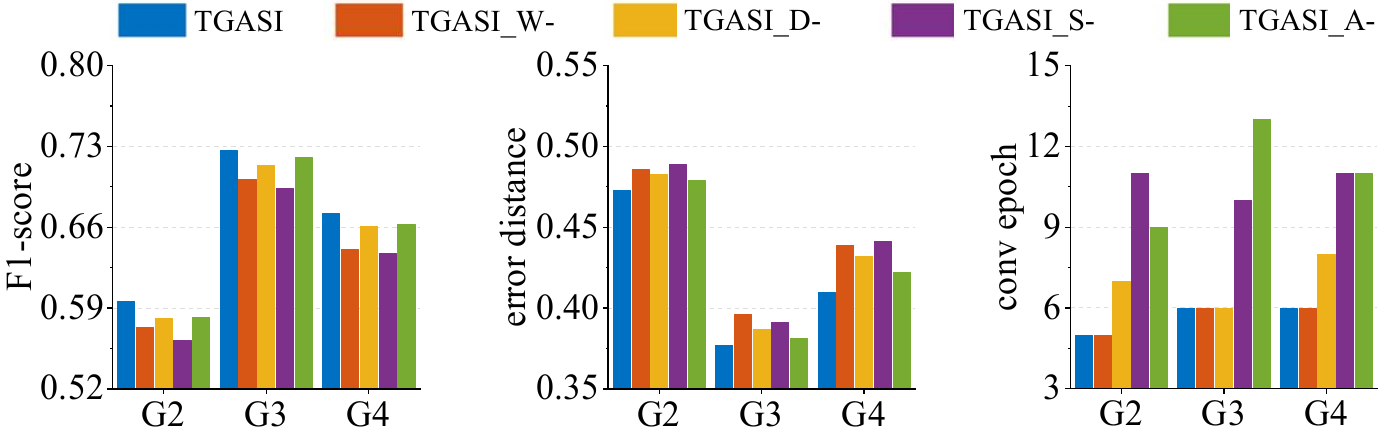}			
	\caption{The performance evaluation of variant models from TGASI in $G_2$ to $G_4$.
	}
	\label{fighaoge}
\end{figure}

\begin{itemize}
    \item $\mathrm{TGASI\_W{-}}$ uses a  zero vector $\left[0\right]^{|v| \times 1}$ to replace the coarse-grained feature $H_1$ shown in Eq.~(\ref{eq6}).
    \item $\mathrm{TGASI\_D{-}}$ uses two zero vectors $\left[0\right]^{|v| \times 1}$ to replace the dynamic features $H_2$ shown in Eq.~(\ref{eq7}) and $H_3$ shown in Eq.~(\ref{eq8}), respectively.
    \item $\mathrm{TGASI\_S{-}}$ uses a zero matrix $\left[0\right]^{|v| \times \sqrt{|V|}}$ to replace the static topology features $H_G$ shown in Eq.~(\ref{eq10}).
    \item $\mathrm{TGASI\_A{-}}$ replaces the attention mechanism in Eqs. (\ref{eq17})-(\ref{eq20}) with $\mathrm{softmax}(\frac{1}{2}( \overset{\rightarrow}{h}_{j}+\overset{\leftarrow}{h}_{j}) )$.
\end{itemize}

We set the parameters of four variant models as the same as those in Sec.~\ref{Optimizations} to guarantee the rigor of the experiments.
Due to the limited space,
we only present the experiment results in $G_2$-$G_4$.
As can be seen from Fig.~\ref{fighaoge} that  it will lead to a performance decrease or a delay in the convergence speed of model training no matter removing any critical modules.

\subsection{Scalability}
The scalability of the model can evaluate whether the model can be applied in a variety of scenarios, which can reflect the strength of its transferability and robustness. The decoder we designed mainly includes Bi-GRU and attention mechanism, which can be used in the inductive learning task.
We demonstrate the inductive learning capabilities of TGASI from two perspectives, including social networks and propagation models.
More specifically, when we have trained the TGASI model using the snapshot set $\{ \mathcal{G}_{s_j}=(V, E,Y_{s_j}) \mid j=1,~2,~3,... \}$ based on a social network $G_1$=$(V,E)$, we further apply the trained TGASI to the other snapshot set $\{ \mathcal{G'}_{s_{j'}}=(V{'}, E{'},Y_{s_{j'}}) \mid j'=1,~2,~3,... \}$ based on the other social network $G_2$=$(V',E')$, where $V$$\neq$$V'$ and $E$$\neq$$E'$, and $Y_{s_{j'}}$ and $Y_{s_j}$ are generated by different propagation models.
Due to the better performance in $G_3$, we use the trained TGASI model in $G_3$ to conduct inductive learning.

\begin{figure}[htb]
	\centering   
	\includegraphics[width=3.25in]{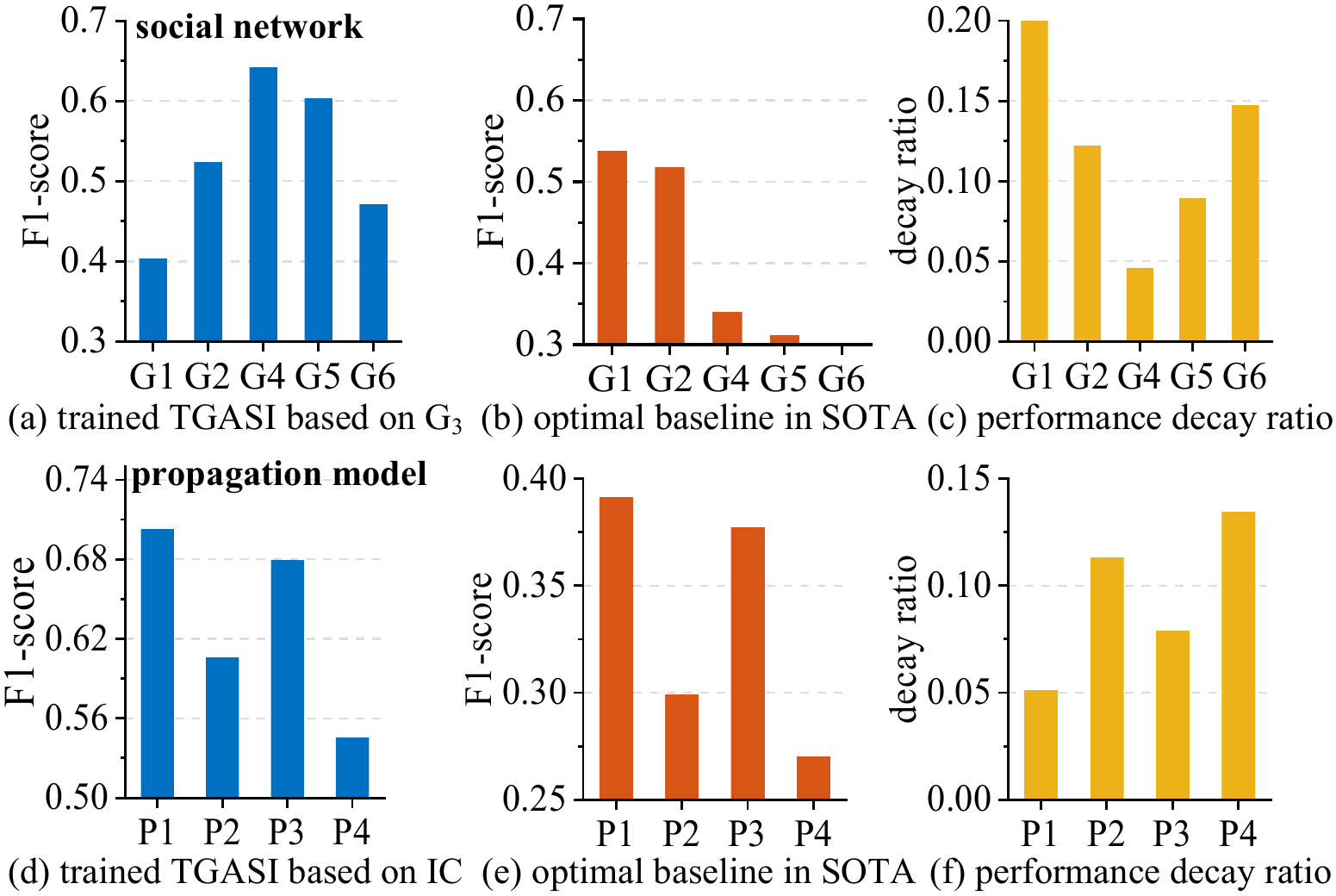}			
	\caption{The performance evaluation of TGASI on the inductive learning task. (a)-(c) is the inductive learning on different social networks based on the IC model, and (d)-(f) is on different propagation models in $G_3$. $P1$, $P2$, $P3$, and $P4$ are denoted as homogeneous SI, homogeneous SIR, heterogeneous SI, and heterogeneous SIR.  
	}
	\label{fig_induc}
\end{figure}

In the inductive learning of social networks, we replace the original test set of $G_3$ with another batch of snapshot sets generated by $G_1$, $G_2$, $G_4$, $G_5$, and $G_6$. One  issue that requires improvement in the future is the variation in the dimensionality of $H_G$ caused by network scale, which only is addressed by using the original topological embedding here. In order to rigorously evaluate TGASI's performance in inductive learning, we also compare the inductive results with the native results of SOTA methods and TGASI on native datasets or propagation models. From figs.~\ref{fig_induc}(a)-(b), the inductive learning ability of TGASI is similar to or better than the original detection performance of the optimal SOTA method.
What's more, it can be seen from Fig.~\ref{fig_induc}(c) that the inductive performance in $G_4$ or $G_5$ is stronger than that in  $G_1$ or $G_6$.
We conclude that the decreased extent of inductive performance  is proportional to the deviation from the network scale corresponding to the trained baseline model.

In the inductive learning of propagation models, we use various propagation models, including the heterogeneous SI with infection rate $I$$\sim$$U(0.05,0.15)$, the heterogeneous SIR with recovery rate $R$$\sim$$U(0,0.05)$, the homogeneous SI with $I=0.1$, and the homogeneous SIR with $R=0.02$, to evaluate the source detection performance of TGASI in different propagation models. 
Figs.~\ref{fig_induc}(d)-(e) show that the inductive learning ability of TGASI based on the propagation model is overall higher than 0.5 on the F-score metric, 
and is better than the original performance of all SOTA methods.
Moreover, the performance of TGASI on the homogeneous model is better than that of the heterogeneous model because it is easier to accurately learn the influence information $\bm{\mathcal{W}}$ in the homogeneous models. 
And the performance  of TGASI on the SI model is better than that of the SIR model because 
the interactive behavior and the temporal characteristic  in the SIR model are more complicated.

\section{Conclusion}\label{sec05}

In this paper, we pay attention to the source localization problem by considering the heterogeneous behavior diversity in time-varying infection scenarios, and we design a sequence-to-sequence framework TGASI with the transferability to share trained TGASI to different propagation models and social networks.
Comprehensive experiments demonstrate the necessity of designed modules, i.e., a coarse-grained-based source probability feature, dynamic infection features, static topology features, a temporal attention mechanism, and a unique loss function for the localization task in TGASI. 
It's worth mentioning that TGASI is the first source localization framework that considers scalability based inductive learning. 
So a trained TGASI model is suitable for other social networks and propagation models.
And model transferability experiments demonstrate that the inductive detection performance of TGASI is overall superior to the original detection performance of SOTA methods. 
However, we also analyze that the source detection performance of TGASI could deteriorate when it transfers to some scenarios.
So we are motivated to propose a novel localization framework in the future that can efficiently handle scalability under various scenarios.





\clearpage
\newpage

\section*{Acknowledgments}
This work was supported by the National Science Fund for
Distinguished Young Scholars (No. 62025602), the National Natural
Science Foundation of China (Nos. 61976181,  62261136549, U22B2036),
Technology-Scientific and Technological Innovation Team of
Shaanxi Province (No. 2020TD-013), Fok Ying-Tong Education Foundation, China (No. 171105), Fundamental Research Funds for the Central Universities (No. D5000211001), and Tencent Foundation and XPLORER PRIZE.

\bibliographystyle{named}

\begin{thebibliography}{}

\bibitem[\protect\citeauthoryear{Agarwal \bgroup \em et al.\egroup
  }{2021}]{agarwal:2021transfer-inductive}
Nidhi Agarwal, Akanksha Sondhi, Khyati Chopra, and Ghanapriya Singh.
\newblock Transfer learning: Survey and classification.
\newblock {\em Smart innovations in communication and computational sciences},
  pages 145--155, 2021.

\bibitem[\protect\citeauthoryear{Baesens \bgroup \em et al.\egroup
  }{2015}]{baesens:2015fraud}
Bart Baesens, Veronique Van~Vlasselaer, and Wouter Verbeke.
\newblock {\em Fraud analytics using descriptive, predictive, and social
  network techniques: a guide to data science for fraud detection}.
\newblock John Wiley \& Sons, 2015.

\bibitem[\protect\citeauthoryear{Chang \bgroup \em et al.\egroup
  }{2018}]{chang:2018reversible}
Bo~Chang, Lili Meng, Eldad Haber, Lars Ruthotto, David Begert, and Elliot
  Holtham.
\newblock Reversible architectures for arbitrarily deep residual neural
  networks.
\newblock In {\em Proceedings of the AAAI conference on artificial
  intelligence}, volume~32, 2018.

\bibitem[\protect\citeauthoryear{Chung \bgroup \em et al.\egroup
  }{2014}]{chung:2014empiricalGRU}
Junyoung Chung, Caglar Gulcehre, Kyunghyun Cho, and Yoshua Bengio.
\newblock Empirical evaluation of gated recurrent neural networks on sequence
  modeling.
\newblock In {\em NIPS 2014 Workshop on Deep Learning}, 12 2014.

\bibitem[\protect\citeauthoryear{Diepold \bgroup \em et al.\egroup
  }{2022}]{diepold:snapshots2022data}
Klaus Diepold, Sven Gronauer, Matthias Kissel, and Mohamed~Ali Tnani.
\newblock Data science for the environment.
\newblock {\em Trend Reports: Seminar on Machine Intelligence}, 2022.

\bibitem[\protect\citeauthoryear{Dong \bgroup \em et al.\egroup
  }{2019}]{dong:cikm}
Ming Dong, Bolong Zheng, Nguyen Quoc Viet~Hung, Han Su, and Guohui Li.
\newblock Multiple rumor source detection with graph convolutional networks.
\newblock In {\em Proceedings of the 28th ACM International Conference on
  Information and Knowledge Management}, pages 569--578, 2019.

\bibitem[\protect\citeauthoryear{Dong \bgroup \em et al.\egroup
  }{2022}]{dong:2022trans}
Ming Dong, Bolong Zheng, Guohui Li, Chenliang Li, Kai Zheng, and Xiaofang Zhou.
\newblock Wavefront-based multiple rumor sources identification by multi-task
  learning.
\newblock {\em IEEE Transactions on Emerging Topics in Computational
  Intelligence}, 2022.

\bibitem[\protect\citeauthoryear{Hochreiter and
  Schmidhuber}{1997}]{hochreiter:1997longLSTM}
Sepp Hochreiter and Jurgen Schmidhuber.
\newblock Long short-term memory.
\newblock {\em Neural computation}, 9(8):1735--1780, 1997.

\bibitem[\protect\citeauthoryear{Jacobsen \bgroup \em et al.\egroup
  }{2018}]{jacobsen:2018revnet}
Jorn-Henrik Jacobsen, Arnold Smeulders, and Edouard Oyallon.
\newblock i-revnet: Deep invertible networks.
\newblock {\em arXiv preprint arXiv:1802.07088}, 2018.

\bibitem[\protect\citeauthoryear{Jiang \bgroup \em et al.\egroup
  }{2016}]{jiang:2016identifying}
Jiaojiao Jiang, Sheng Wen, Shui Yu, Yang Xiang, and Wanlei Zhou.
\newblock Identifying propagation sources in networks: State-of-the-art and
  comparative studies.
\newblock {\em IEEE Communications Surveys \& Tutorials}, 19(1):465--481, 2016.

\bibitem[\protect\citeauthoryear{Jin and Wu}{2021}]{jin:2021schemes}
Rong Jin and Weili Wu.
\newblock Schemes of propagation models and source estimators for rumor source
  detection in online social networks: A short survey of a decade of research.
\newblock {\em Discrete Mathematics, Algorithms and Applications},
  13(04):2130002, 2021.

\bibitem[\protect\citeauthoryear{Jin \bgroup \em et al.\egroup
  }{2017}]{jin:2017multimodal}
Zhiwei Jin, Juan Cao, Han Guo, Yongdong Zhang, and Jiebo Luo.
\newblock Multimodal fusion with recurrent neural networks for rumor detection
  on microblogs.
\newblock In {\em Proceedings of the 25th ACM international conference on
  Multimedia}, pages 795--816, 2017.

\bibitem[\protect\citeauthoryear{Khoo \bgroup \em et al.\egroup
  }{2020}]{khoo:2020interpretable}
Ling Min~Serena Khoo, Hai~Leong Chieu, Zhong Qian, and Jing Jiang.
\newblock Interpretable rumor detection in microblogs by attending to user
  interactions.
\newblock In {\em Proceedings of the AAAI conference on artificial
  intelligence}, volume~34, pages 8783--8790, 2020.

\bibitem[\protect\citeauthoryear{Kou and Gray}{2017}]{kou:2017supporting}
Yubo Kou and Colin~M Gray.
\newblock Supporting distributed critique through interpretation and
  sense-making in an online creative community.
\newblock {\em Proceedings of the ACM on Human-Computer Interaction},
  1(CSCW):1--18, 2017.

\bibitem[\protect\citeauthoryear{Lao \bgroup \em et al.\egroup
  }{2021}]{lao:2021rumor}
An~Lao, Chongyang Shi, and Yayi Yang.
\newblock Rumor detection with field of linear and non-linear propagation.
\newblock In {\em Proceedings of the Web Conference}, pages 3178--3187, 2021.

\bibitem[\protect\citeauthoryear{Li \bgroup \em et al.\egroup
  }{2021}]{li:2021JBI}
Liang Li, Jianye Zhou, Yuewen Jiang, and Biqing Huang.
\newblock Propagation source identification of infectious diseases with graph
  convolutional networks.
\newblock {\em Journal of biomedical informatics}, 116:103720, 2021.

\bibitem[\protect\citeauthoryear{Ling \bgroup \em et al.\egroup
  }{2022}]{ling:2022kdd}
Chen Ling, Junji Jiang, Junxiang Wang, and Zhao Liang.
\newblock Source localization of graph diffusion via variational autoencoders
  for graph inverse problems.
\newblock In {\em Proceedings of the 28th ACM SIGKDD Conference on Knowledge
  Discovery and Data Mining}, pages 1010--1020, 2022.

\bibitem[\protect\citeauthoryear{Michalski}{1983}]{michalski:1983theory}
Ryszard~S Michalski.
\newblock A theory and methodology of inductive learning.
\newblock In {\em Machine learning}, pages 83--134. Elsevier, 1983.

\bibitem[\protect\citeauthoryear{Paluch \bgroup \em et al.\egroup
  }{2020}]{paluch:2020optimizing}
Robert Paluch, Lukasz~G Gajewski, Janusz~A Holyst, and Boleslaw~K Szymanski.
\newblock Optimizing sensors placement in complex networks for localization of
  hidden signal source: A review.
\newblock {\em Future Generation Computer Systems}, 112:1070--1092, 2020.

\bibitem[\protect\citeauthoryear{Qian \bgroup \em et al.\egroup
  }{2018}]{qian:2018neural}
Feng Qian, Chengyue Gong, Karishma Sharma, and Yan Liu.
\newblock Neural user response generator: Fake news detection with collective
  user intelligence.
\newblock In {\em Proceedings of the International Joint Conference on
  Artificial Intelligence}, volume~18, pages 3834--3840, 2018.

\bibitem[\protect\citeauthoryear{Salamat \bgroup \em et al.\egroup
  }{2020}]{salamat:2020balnode2vec}
Amirreza Salamat, Xiao Luo, and Ali Jafari.
\newblock Balnode2vec: Balanced random walk based versatile feature learning
  for networks.
\newblock In {\em 2020 International Joint Conference on Neural Networks
  (IJCNN)}, pages 1--8. IEEE, 2020.

\bibitem[\protect\citeauthoryear{Shah \bgroup \em et al.\egroup
  }{2020}]{shah:2020patient0}
Chintan Shah, Nima Dehmamy, Nicola Perra, and Matteo Chinazzi.
\newblock Finding patient zero: Learning contagion source with graph neural
  networks.
\newblock {\em arXiv preprint arXiv:2006.11913}, 2020.

\bibitem[\protect\citeauthoryear{Shelke and Attar}{2019}]{shelke:2019source}
Sushila Shelke and Vahida Attar.
\newblock Source detection of rumor in social network--a review.
\newblock {\em Online Social Networks and Media}, 9:30--42, 2019.

\bibitem[\protect\citeauthoryear{Sokolova \bgroup \em et al.\egroup
  }{2006}]{sokolova:2006F1}
Marina Sokolova, Nathalie Japkowicz, and Stan Szpakowicz.
\newblock Beyond accuracy, f-score and roc: A family of discriminant measures
  for performance evaluation.
\newblock In {\em Australasian Joint Conference on Artificial Intelligence},
  pages 1015--1021, 2006.

\bibitem[\protect\citeauthoryear{Song \bgroup \em et al.\egroup
  }{2021}]{song:2021multimodal}
Chenguang Song, Nianwen Ning, Yunlei Zhang, and Bin Wu.
\newblock A multimodal fake news detection model based on crossmodal attention
  residual and multichannel convolutional neural networks.
\newblock {\em Information Processing \& Management}, 58(1):102437, 2021.

\bibitem[\protect\citeauthoryear{Sun \bgroup \em et al.\egroup
  }{2021}]{sun:2021multichanel}
Kuangchi Sun, Zhenfeng Huang, Hanling Mao, Aisong Qin, Xinxin Li, Weili Tang,
  and Jianbin Xiong.
\newblock Multi-scale cluster-graph convolution network with multi-channel
  residual network for intelligent fault diagnosis.
\newblock {\em IEEE Transactions on Instrumentation and Measurement}, 71:1--12,
  2021.

\bibitem[\protect\citeauthoryear{Sutskever \bgroup \em et al.\egroup
  }{2014}]{sutskever:2014sequence}
Ilya Sutskever, Oriol Vinyals, and Quoc~V Le.
\newblock Sequence to sequence learning with neural networks.
\newblock {\em Advances in neural information processing systems}, 27, 2014.

\bibitem[\protect\citeauthoryear{Velickovic \bgroup \em et al.\egroup
  }{2018}]{velivckovic:2018GAT}
Petar Velickovic, Guillem Cucurull, Arantxa Casanova, Adriana Romero, Pietro
  Lio, and Yoshua Bengio.
\newblock Graph attention networks.
\newblock {\em International Conference on Learning Representations}, 2018.

\bibitem[\protect\citeauthoryear{Ventola}{2014}]{ventola:2014social}
C~Lee Ventola.
\newblock Social media and health care professionals: benefits, risks, and best
  practices.
\newblock {\em Pharmacy and therapeutics}, 39(7):491, 2014.

\bibitem[\protect\citeauthoryear{Wang \bgroup \em et al.\egroup
  }{2017}]{wang:2017multipleAAAI}
Zheng Wang, Chaokun Wang, Jisheng Pei, and Xiaojun Ye.
\newblock Multiple source detection without knowing the underlying propagation
  model.
\newblock In {\em Proceedings of the AAAI Conference on Artificial
  Intelligence}, pages 217--223, San Francisco, CA, 2017. PALO ALTO, CA 94303
  USA.

\bibitem[\protect\citeauthoryear{Wang \bgroup \em et al.\egroup
  }{2022a}]{wang:2022www}
Junxiang Wang, Junji Jiang, and Liang Zhao.
\newblock An invertible graph diffusion neural network for source localization.
\newblock In {\em Proceedings of the ACM Web Conference}, pages 1058--1069,
  2022.

\bibitem[\protect\citeauthoryear{Wang \bgroup \em et al.\egroup
  }{2022b}]{wang:2022rapid}
Zhen Wang, Dongpeng Hou, Chao Gao, Jiajin Huang, and Qi~Xuan.
\newblock A rapid source localization method in the early stage of large-scale
  network propagation.
\newblock In {\em Proceedings of the ACM Web Conference}, pages 1372--1380,
  2022.

\bibitem[\protect\citeauthoryear{Welling and Kipf}{2016}]{welling:2016GCNICLR}
Max Welling and Thomas~N Kipf.
\newblock Semi-supervised classification with graph convolutional networks.
\newblock In {\em International Conference on Learning Representations (ICLR
  2017)}, 2016.

\bibitem[\protect\citeauthoryear{Xia \bgroup \em et al.\egroup
  }{2021}]{xia:WSDM2021deepis}
Wenwen Xia, Yuchen Li, Jun Wu, and Shenghong Li.
\newblock Deepis: Susceptibility estimation on social networks.
\newblock In {\em Proceedings of the 14th ACM International Conference on Web
  Search and Data Mining}, pages 761--769, 2021.

\bibitem[\protect\citeauthoryear{Yang \bgroup \em et al.\egroup
  }{2021}]{yang:2021rumor}
Xiaoyu Yang, Yuefei Lyu, Tian Tian, Yifei Liu, Yudong Liu, and Xi~Zhang.
\newblock Rumor detection on social media with graph structured adversarial
  learning.
\newblock In {\em Proceedings of the twenty-ninth international conference on
  international joint conferences on artificial intelligence}, pages
  1417--1423, 2021.

\bibitem[\protect\citeauthoryear{Zhang \bgroup \em et al.\egroup
  }{2019}]{zhang:2019trafficTransfer}
Zhengchao Zhang, Meng Li, Xi~Lin, Yinhai Wang, and Fang He.
\newblock Multistep speed prediction on traffic networks: A deep learning
  approach considering spatio-temporal dependencies.
\newblock {\em Transportation research part C: emerging technologies},
  105:297--322, 2019.

\bibitem[\protect\citeauthoryear{Zhu \bgroup \em et al.\egroup
  }{2022}]{zhu:2022learnableDynamic}
Yuecai Zhu, Fuyuan Lyu, Chengming Hu, Xi~Chen, and Xue Liu.
\newblock Learnable encoder-decoder architecture for dynamic graph: A survey.
\newblock {\em arXiv preprint arXiv:2203.10480}, 2022.

\end{thebibliography}

\end{document}